\documentstyle[11pt,epsfig]{article}
\title{\bf Can local bulk effects explain the galactic dark matter?}
\author{Malihe Heydari-Fard $^{1,2}$\thanks{email:
m.heydarifard@mail.sbu.ac.ir} and Hamid R. Sepangi
\thanks{email: hr-sepangi@sbu.ac.ir}
\\ {\small $^1$Department
of Physics, Shahid Beheshti University, Evin, Tehran 19839, Iran}
\\ {\small $^2$Department of Physics, The University of Qom, Qom
37185-359, Iran}}
\begin{document}
\maketitle 
\begin{abstract}
We obtain the virial theorem within the context of a brane-world
model without mirror symmetry or any form of junction condition.
Taking a constant curvature bulk (neglecting non-local bulk
effects), the local bulk effects generate a geometrical mass,
contributing to the gravitational energy which may be used to
explain the virial mass discrepancy in clusters of galaxies. We
fix the parameter of this model in agreement with observational
data.
\vspace{5mm}
\end{abstract}
\section{Introduction}
The question of dark matter is one of the most intriguing aspects
of the present-day observational astrophysics and cosmology.
Today, its existence at the galactic and extra-galactic scales is
well established and is based on two important observations,
namely the mass discrepancy in clusters of galaxies and the
behavior of the galactic rotation curves \cite{book}. Galaxy
clusters are the largest virialized structures in the universe,
and their mass content is supposed to be representative of the
universe as a whole. The total mass of galaxy clusters ranges from
$10^{13} M_{\odot}$ for groups up to a few $10^{15} M_{\odot}$ for
very rich systems. Cluster masses can be determined in a variety
of ways. The application of the virial theorem to positions and
velocities of cluster member galaxies is the oldest method of
cluster mass determination \cite{1}. More recent methods are based
on the dynamical analysis of hot X-ray emitting gas \cite{2} and
on the gravitational lensing of background galaxies \cite{3}. The
mass determined from such dynamical means is always found to be in
excess of that which can be attributed to the visible matter. This
is known as the ``missing mass problem.'' The existence of dark
matter was not firmly established until the measurement of the
rotational velocity of stars and gases orbiting at a distance $r$
from the galactic center could be done with reasonable accuracy.
Observations show that the rotational velocity increases near the
center of a galaxy and approaches a nearly constant value with
increasing distance from the center. The discrepancy between the
observed rotational velocity curves and the theoretical prediction
from Newtonian mechanics is known as the ``galactic rotation
curves problem.'' This discrepancy is explained by postulating
that every galaxy and cluster of galaxies is embedded in a halo
made up of some dark (invisible) matter.

The idea that our familiar 4-dimensional ($4D$) space-time is a
hypersurface (brane) embedded in a 5D bulk
\cite{Nima,Randall,Dvali} has been experiencing a phenomenal
interest during the last decade. According to this brane-world
scenario, all matter and gauge interactions reside on the brane,
while gravity can propagate in the whole 5D space-time. Several
brane-world cosmologies have been proposed in the context of the
Randall-Sundrum (RS) formulations \cite{Randall}, defined in a 5D
anti-de Sitter space-time (AdS$_5$). The dynamics of these models
feature boundary terms in the action and sometimes mirror
symmetry, such that bulk gravitational waves interfere with the
brane-world motion. This usually comes together with junction
conditions producing an algebraic relationship between the
extrinsic curvature and the confined matter \cite{israel,Battye}.
The consequence is that the Friedman equation acquires an
additional term which is proportional to the square of energy
density of the confined matter field \cite{Cline,Binetruy}. This
term was initially considered as a possible solution to the
accelerated expansion of the universe. However, soon it was
realized to be incompatible with the big bang nucleosynthesis,
requiring additional fixes \cite{Binetruy}.

Brane-world scenarios under more general conditions and still
compatible with the brane-world program have also been rather
extensively studied over the past decade where it has been shown
that it is possible to find a richer set of cosmological solutions
in accordance with the current observations \cite{Maia}. Under
these conditions, without using $Z_2$ symmetry or without
postulating any junction condition, Friedman equation is modified
by a geometrical term which is defined in terms of the extrinsic
curvature, leading to a geometrical interpretation for dark energy
\cite{maia}. There has also been arguments concerning the
uniqueness of the junction conditions. Indeed, other forms of
junction conditions exist, so that different conditions may lead
to different physical results \cite{Battye}. Furthermore, these
conditions cannot be used when more than one non-compact extra
dimension is involved. Against this background, an interesting
higher-dimensional model was introduced in \cite{Rubakov} where
particles are trapped on a 4D hypersurface by the action of a
confining potential. The dynamics of test particles confined to a
brane by the action of such potential at the classical and quantum
levels were studied in \cite{shahram}. In \cite{fard}, the same
brane-world model was studied, offering a geometrical explanation
for the accelerated expansion of the universe. A geometrical
explanation for the generalized Chaplygin gas was considered in
\cite{gas} along the same lines. We have also studied exact
solutions of the vacuum field equations on the brane for two
interesting cases. The first solution can be used to explain the
galaxy rotation curves without assuming the existence of dark
matter and without having to resort to the Modified Newtonian
Dynamics (MOND) \cite{Milgrom,Moffat}, and the second solution
represents a black hole in an asymptotically de Sitter space-time
\cite{Razmi}.

It has been argued that modified theories of gravity based on the
RS brane-world scenario can explain the observed galactic rotation
curves without introducing any additional hypothesis. In such
models, a spherically symmetric brane-world metric results in a
Schwarzschild mass parameter receiving a new contribution due to
the Weyl tensor which is interpreted as the mass of dark matter
\cite{Mak}. Neglecting the cosmological constant, the dark mass
increases linearly with the radial distance from the galactic
center, explaining the flatness of the galactic rotation curves.
The bending angle of light predicted by brane-world modeles was
found to be much larger compared to the predictions of dark matter
models, the deviation increasing with the degree of compactness
\cite{Cheng}. Using the smallness of the rotational velocity, a
perturbation scheme for reconstructing the metric in a galactic
halo in brane-world models with induced gravity was developed in
\cite{Viznyuk}. It has also been shown that the non-standard model
bulk fields can replace dark matter in explaining structure
formation, as the evolution on the brane becomes similar to that
of the cold dark matter \cite{Pal}. Similar interpretations of
dark matter as bulk effects have also been considered in
\cite{Boehmer,pal}.

As was mentioned above, the galactic dynamics of massive test
particles may be understood without resorting to dark matter in
the framework of $f(R)$ modified theories of gravity \cite{S}.
Recently, it has been argued that the virial theorem mass
discrepancy in clusters of galaxies can be accounted for within
the context of the RS model and $f(R)$ modified theories of
gravity \cite{Harko,f(R)}. It would therefore be of interest to
study the latter in a brane-world model with a constant curvature
bulk without using the $Z_2$ symmetry or without postulating any
form of junction conditions \cite{Maia,maia}. In so doing the
gravitational field equations on the brane are modified by a local
extra term, $Q_{\mu\nu}$. Using the collisionless Boltzmann
equation, we obtain the virial theorem which is modified by an
extra term that may be used to explain the virial mass discrepancy
in clusters of galaxies.
\section{Field equations for the brane-world}
The embedding of the brane-world in the bulk plays an essential
role in the covariant formulation of the brane-world gravity,
since it tells us how the Einstein-Hilbert dynamics of the bulk is
transferred to the brane-world. However, there are many different
ways to embed a manifold into another, classified as local,
global, isometric, conformal, rigid, deformable, analytic or
differentiable. The choice of one or other depends on what the
embedded manifold is supposed to do.

Generally, there are three basic postulates in the geometrical
approach considered in brane-world scenarios, that is, the
confinement of the standard gauge interactions to the brane, the
existence of quantum gravity in the bulk and finally, the
embedding of the brane-world. All other model dependent properties
such as warped metric, mirror symmetries, radion or extra scalar
fields, fine tuning parameters like the tension of the brane and
the choice of a junction condition are left out as much as
possible in our calculations \cite{Maia}.

In the following we extend this embedding to any constant
curvature space, including anti-de Sitter $AdS_{5}$, de Sitter
$dS_{5}$, and the flat $M_{5}$ cases, with the bulk Riemann tensor
written as
\begin{eqnarray}
{\cal R}_{ABCD}=k_{*}\left({\cal G}_{AC}{\cal G}_{BD}-{\cal
G}_{AD}{\cal G}_{BC}\right),\label{1}
\end{eqnarray}
where ${\cal G}_{AB}$ is the bulk metric. In the normal Gaussian
frame defined by the embedded space-time, it may be decomposed as
\begin{eqnarray}
{\cal G}_{AB}=\left( \!\!\!
\begin{array}{cc}
g_{\mu \nu } & 0 \\
0 & g_{55}
\end{array}
\!\!\!\right) ,\hspace{.5 cm}g_{55}=\varepsilon=\pm1,
\end{eqnarray}
where $k_{*}$ denotes the bulk constant curvature. In the flat
case $k_{*}=0$ and we may write
$k_{*}=\varepsilon\frac{\Lambda^{(b)}}{6}$, where $\Lambda^{(b)}$
is the bulk cosmological constant. We note that for a $dS_{5}$
bulk $\varepsilon=+1$ and for a $AdS_{5}$ bulk $\varepsilon=-1$.
With this assumption the Gauss-Codazzi equations \cite{Book}
\begin{eqnarray}
R_{\alpha\beta\gamma\delta}
=\frac{1}{\varepsilon}(K_{\alpha\gamma}K_{\beta\delta}-K_{\alpha\delta}K_{\beta\gamma})
+{\cal R}_{ABCD} {\cal Z}_{,\alpha}^{A}{\cal Z}^{B}_{,\beta}{\cal
Z}^{C}_{,\gamma}{\cal Z}^{D}_{,\delta},\label{2}
\end{eqnarray}
\begin{eqnarray}
K_{\alpha[\beta;\gamma]} = {\cal R}_{ABCD} {\cal
Z}_{,\alpha}^{A}{\cal N}^{B}{\cal Z}^{C}_{,\beta}{\cal
Z}^{D}_{,\gamma},\label{2}
\end{eqnarray}
which are the integrability conditions for the embedding
equations, reduce to
\begin{eqnarray}
R_{\alpha\beta\gamma\delta}
=\frac{1}{\varepsilon}(K_{\alpha\gamma}K_{\beta\delta}-K_{\alpha\delta}K_{\beta\gamma})
+ k_{*}
(g_{\alpha\gamma}g_{\beta\delta}-g_{\alpha\delta}g_{\beta\gamma}),\label{3}
\end{eqnarray}
\begin{eqnarray}
K_{\alpha[\beta;\gamma]} = 0,\label{4}
\end{eqnarray}
where $K_{\mu\nu}$ denotes the extrinsic curvature. From the
contractions of equation (\ref{3}) we can obtain the Ricci scalar
$R$ of the brane metric, in terms of the Ricci scalar of the bulk
$\cal{R}$. After adding a confined matter Lagrangian and a four
dimensional cosmological constant $\Lambda$, the effective
Lagrangian compatible with the embedding reads, for details see
\cite{maiaa}
\begin{eqnarray}
{\cal L}_{eff}=R
\sqrt{-g}-\frac{1}{\varepsilon}\left(K^{\mu\nu}K_{\mu\nu}+K^2\right)\sqrt{-g}+{\cal
R}\sqrt{-g}-\frac{2}{\varepsilon}\frac{dK}{dy}\sqrt{-g}+\Lambda\sqrt{-g}-{\cal
L}_{m},\label{5}
\end{eqnarray}
where $K=g^{\mu\nu}K_{\mu\nu}$ is the mean curvature of the
brane-world, $y$ denotes the fifth coordinate and ${\cal R}
=-20k_{*}$ as derived from equation (\ref{1}). The total
derivative term with respect to $y$ can be eliminated provided the
motion of the brane-world occurs between two fixed minimal
boundary hypersurfaces where $K = 0$ \cite{Maia}. Similar to the
RSII model, such boundaries can be moved away so that all boundary
generated bulk waves are eliminated. Variation of the action with
respect to $g_{\mu\nu}$ gives the dynamical equation compatible
with the embedding and with the confined matter hypotheses
\begin{eqnarray}
G_{\mu\nu}=8\pi G\tau_{\mu\nu}- \lambda g_{\mu\nu}+
Q_{\mu\nu},\label{6}
\end{eqnarray}
where
\begin{eqnarray}
Q_{\mu\nu}=\frac{1}{\varepsilon}\left[K^{\rho}_{\,\,\,\,\mu
}K_{\rho\nu}-KK_{\mu\nu}-\frac{1}{2}
\left(K_{\alpha\beta}K^{\alpha\beta}-K^2\right)g_{\mu\nu}\right].\label{7}
\end{eqnarray}
Here, $\tau_{\mu\nu}$ is the confined matter energy-momentum
tensor on the brane and $\lambda=-3k_{*}+\Lambda$. As can be seen
from definition of $Q_{\mu\nu}$,  it is independently a conserved
quantity, that is
\begin{eqnarray}
Q^{\mu\nu}_{\,\,\,\,;\mu}=0.\label{8}
\end{eqnarray}
so that there is no non-gravitational exchange of energy between
this geometrical correction and the confined matter. Such an
aspect has one important consequence, that is, if $Q_{\mu\nu}$ is
to be related to  dark energy, it does not exchange energy with
ordinary matter, much the same as in coupled quintessence models
\cite{W}.

In order to solve the Codazzi equation (\ref{4}), we choose the
static spherically symmetric metric on the brane in the form
\begin{eqnarray}
ds^2=-e^{\mu(r)}dt^2+e^{\nu(r)}dr^2+r^2\left(d\theta^2+\sin^2\theta
d\varphi^2\right).\label{9}
\end{eqnarray}
Then the York relation
\begin{eqnarray}
K_{\mu \nu }=-\frac{1}{2}\frac{\partial
g_{\mu\nu}}{\partial\xi},\label{10}
\end{eqnarray}
shows that in a diagonal metric, $K_{\mu\nu }$ are diagonal. Here,
$\xi$ is a small parameter orthogonal to the brane that
parameterizes the extra noncompact dimension. Now, separating the
spatial components, the Codazzi equation reduces to
\begin{equation}
K_{\mu\nu ,\sigma}-K_{\nu\rho }\Gamma^{\rho}_{\mu\sigma}=
K_{\mu\sigma ,\nu}-K_{\sigma\rho
}\Gamma^{\rho}_{\mu\nu},\label{12}
\end{equation}
\begin{eqnarray}
K_{00,1}-\left(\frac{\mu^{'}}{2}\right)K_{00}=-\left(\frac{\mu^{'}e^\mu}{2e^\nu}\right)
K_{11},\label{13}
\end{eqnarray}
\begin{eqnarray}
K_{22,1}-\left(\frac{1}{r}\right)K_{22}=\left({re^{-\nu}}\right)
K_{11}.\label{14}
\end{eqnarray}
The first equation gives
$K_{00,\sigma}=K_{11,\sigma}=K_{22,\sigma}=K_{33,\sigma}=0$ for
$\sigma=0,3$. Repeating the same procedure for $\sigma=2$, we
obtain $K_{00,\sigma}=K_{11,\sigma}=K_{22,\sigma}=0$. This shows
that $K_{11}$ depends only on the variable $r$. Assuming
$K_{11}=\alpha e^{\nu(r)}$ and using equations (\ref{13}) and
(\ref{14}), one finds
\begin{eqnarray}
K_{00}(r)=-\alpha e^{\mu(r)}+ce^{\mu(r)/2},\label{15}
\end{eqnarray}
\begin{eqnarray}
K_{22}(r)=\alpha r^2+\beta r.\label{16}
\end{eqnarray}
Taking $\mu,\nu=3$ in the first equation we obtain
\begin{eqnarray}
K_{33,1}-\left(\frac{1}{r}\right)K_{33}=\left({e^{-\nu}r\sin^2{\theta}}
\right)K_{11}=\alpha r\sin^2{\theta},\label{17}
\end{eqnarray}
\begin{eqnarray}
K_{33,2}-\left(\cot{\theta}\right)K_{33}=\left
(\sin{\theta}\cos{\theta}\right)K_{22}.\label{18}
\end{eqnarray}
Using equations (\ref{15}), (\ref{16}) and (\ref{17}) then results
in
\begin{eqnarray}
K_{33}(r,\theta)=\alpha r^2\sin^2{\theta}+r\beta
\sin^2{\theta}+rc_{1}\sin{\theta}.\label{19}
\end{eqnarray}
Now, use of equation (\ref{7}) leads to the components of
$Q_{\mu\nu}$
\begin{eqnarray}\label{EQ}
Q_{00}&=&-\frac{g_{00}}{\varepsilon r^2}
\left[3\alpha^2r^2+4\alpha\beta
r+\beta^2+\frac{c_1}{\sin{\theta}}(2\alpha
r+\beta) \right],\nonumber\\
Q_{11}&=&- \frac{g_{11}}{\varepsilon
r^2}\left[3\alpha^2r^2+4\alpha\beta
r+\beta^2+\frac{c_1}{\sin{\theta}}\left(2\alpha r+\beta-c
re^{-\mu/2}\right)-2ce^{-\mu/2}\left(\alpha r^2+\beta
r\right)\right], \nonumber\\
Q_{22}&=& \frac{g_{22}}{\varepsilon
r}\left[-3\alpha^2r-2\alpha\beta+ce^{-\mu/2}(2\alpha
r+\beta)+\frac{c_1}{\sin{\theta}}
\left(-2\alpha+ce^{-\mu/2}\right)\right],\nonumber\\
Q_{33}&=& \frac{g_{33}}{\varepsilon
r}\left[-3\alpha^2r-2\alpha\beta+ce^{-\mu/2}(2\alpha
r+\beta)\right].
\end{eqnarray}
Since $G_{2}^{2}=G_{3}^{3}$ and thus $Q_{2}^{2}=Q_{3}^{3}$, one
obtains $c_1=0$. Now, using these relations and equation
(\ref{6}), the gravitational field equations become
\begin{eqnarray}
{e^{-\nu}}\left(-\frac{1}{r^2}+\frac{\nu^{'}}{r}\right)+\frac{1}{r^2}=
8\pi
G\rho_{eff}+\lambda+\frac{1}{\varepsilon}\left(3\alpha^2+\frac{4\alpha\beta}
{r}+\frac{\beta^2}{r^2}\right),\label{20}
\end{eqnarray}
\begin{eqnarray}
e^{-\nu}\left(\frac{1}{r^2}+\frac{\mu^{'}}{r}\right)-\frac{1}{r^2}=
8\pi
Gp^{(r)}_{eff}-\lambda+\frac{1}{\varepsilon}\left(-3\alpha^2+2\alpha
ce^{-\mu/2}-\frac{4\alpha\beta}
{r}-\frac{\beta^2}{r^2}+\frac{2\beta
c}{r}e^{-\mu/2}\right),\label{21}
\end{eqnarray}
\begin{eqnarray}
e^{-\nu}\left(\frac{\mu^{'}}{r}-\frac{\nu^{'}}{r}-\frac{\mu^{'}\nu^{'}}{2}
+{\mu^{''}}+\frac{\mu^{'2}}{2}\right)=16\pi
Gp^{(\perp)}_{eff}-2\lambda+\frac{1}{\varepsilon}\left(-6\alpha^2+4\alpha
ce^{-\mu/2}-\frac{4\alpha\beta}{r}+\frac{2\beta
c}{r}e^{-\mu/2}\right),\label{22}
\end{eqnarray}
where a prime represents differentiation with respect to $r$ and
$\varepsilon$ is the signatures of the bulk space,
$\varepsilon=+1$ for the $dS_{5}$ bulk and $\varepsilon=-1$ for
the $AdS_{5}$ bulk. In the above equations we have assumed that
the matter on the brane consists of an anisotropic fluid,
characterized by an effective energy density $\rho_{eff}\neq 0$, a
radial pressure $p^{(r)}_{eff}$ and a tangential pressure
$p^{(\perp)}_{eff}$.

In the next section, we will investigate the influence of the
extra term, $Q_{\mu\nu}$, on the dynamics of the galaxies by
adopting the method introduced in \cite{Harko}.
\section{Dark matter and the role of extrinsic curvature}
In order to obtain the virial theorem for galaxy clusters in our
model, we have to write down the general relativistic Boltzmann
equation governing the evolution of the distribution function $f$.
Galaxies, which are treated as identical and collisionless point
particles, are described by this distribution function. For the
static spherically symmetric metric given by equation (\ref{9}) we
introduce the following frame of orthonormal vectors
\cite{Jackson,tetrad}
\begin{eqnarray}
e^{(0)}_{\rho}=e^{\mu/2}\delta^{0}_{\rho},\hspace{.5
cm}e^{(1)}_{\rho}=e^{\nu/2}\delta^{1}_{\rho},\hspace{.5
cm}e^{(2)}_{\rho}=r\delta^{2}_{\rho},\hspace{.5
cm}e^{(3)}_{\rho}=r\sin\theta\delta^{3}_{\rho},
\end{eqnarray}
where $g^{\mu\nu}e_{\mu}^{(a)}e_{\nu}^{(b)}=\eta^{(a)(b)}$. The
four velocity $v^{\mu}$ of a typical galaxy, satisfying the
condition $v^{\mu}v_{\mu}=-1$, in tetrad components is given by
\begin{eqnarray}
v^{(a)}=v^{\mu}e^{(a)}_{\mu}.
\end{eqnarray}
The relativistic Boltzmann equation in tetrad components is given
by
\begin{eqnarray}
v^{(a)}e^{\rho}_{(a)}\frac{\partial f}{\partial
x^{\rho}}+\gamma^{(a)}_{(b)(c)}v^{(b)}v^{(c)}\frac{\partial
f}{\partial v^{(a)}}=0,\label{23}
\end{eqnarray}
where the distribution function $f=f(x^{\mu},v^{(a)})$,
$a=0,1,2,3$, and
$\gamma^{(a)}_{(b)(c)}=e^{(a)}_{\rho;\sigma}e^{\rho}_{(b)}e^{\sigma}_{(c)}$
are the Ricci rotation coefficients. Thus, equation (\ref{23})
becomes
\begin{eqnarray}
v_{r}\frac{\partial f}{\partial
r}&+&\frac{e^{\nu/2}v_{\theta}}{r}\frac{\partial f}{\partial
\theta}+\frac{e^{\nu/2}v_{\varphi}}{r \sin\theta}\frac{\partial
f}{\partial \varphi}-\left(\frac{v_{t}^{2}}{2}\frac{\partial
\mu}{\partial r}
-\frac{v_{\theta}^{2}+v_{\varphi}^{2}}{r}\right)\frac{\partial
f}{\partial v_{r}}
\nonumber\\
&-&\frac{v_{r}}{r}\left(v_{\theta}\frac{\partial f}{\partial
v_{\theta}}+v_{\varphi}\frac{\partial f}{\partial
v_{\varphi}}\right)-\frac{e^{\nu/2}v_{\varphi}}{r}\cot{\theta}\left(v_{\theta}\frac{\partial
f}{\partial v_{\varphi}}-v_{\varphi}\frac{\partial f}{\partial
v_{\theta}}\right)=0.\label{23new}
\end{eqnarray}
Assuming that the distribution function is only a function of the
radial coordinate $r$, equation (\ref{23new}) reduces to
\begin{eqnarray}
v_{r}\frac{\partial f}{\partial
r}-\left(\frac{v_{t}^{2}}{2}\frac{\partial \mu}{\partial
r}-\frac{v_{\theta}^{2}+v_{\varphi}^{2}}{r}\right)\frac{\partial
f}{\partial v_{r}}-\frac{v_{r}}{r}\left(v_{\theta}\frac{\partial
f}{\partial v_{\theta}}+v_{\varphi}\frac{\partial f}{\partial
v_{\varphi}}\right)-\frac{e^{\nu/2}v_{\varphi}}{r}\cot{\theta}\left(v_{\theta}\frac{\partial
f}{\partial v_{\varphi}}-v_{\varphi}\frac{\partial f}{\partial
v_{\theta}}\right)=0.\label{24}
\end{eqnarray}
Now, taking into account the spherically symmetric nature of the
problem at hand and integrating over the velocity space and
assuming that the distribution function vanishes rapidly as the
velocities tend to $\pm \infty$, one obtains
\begin{eqnarray}
r\frac{\partial}{\partial
r}\left[\rho\langle{v_r^2}\rangle\right]+\frac{1}{2}\rho\left[\langle{v_t^2}\rangle+\langle{v_r^2}\rangle\right]
r\frac{\partial \mu}{\partial
r}-\rho\left[\langle{v_\theta^2}\rangle+\langle{v_\varphi^2}\rangle-2\langle{v_r^2}\rangle\right]=0.\label{25}
\end{eqnarray}
Multiplying equation (\ref{25}) by $4\pi r^2$ and integrating over
the cluster leads to
\begin{eqnarray}
-\int^{R}_{0}4\pi\rho\left[\langle{v_r^2}\rangle+\langle{v_\theta^2}\rangle+\langle{v_\varphi^2}\rangle\right]r^2
dr+\frac{1}{2}\int^{R}_{0}4\pi r^3\rho
\left[\langle{v_t^2}\rangle+\langle{v_r^2}\rangle\right]\frac{\partial
\mu}{\partial r}dr=0.\label{26}
\end{eqnarray}
Now, using the total kinetic energy of the galaxies
\begin{eqnarray}
k=\int^{R}_{0}2\pi\rho\left[\langle{v_r^2}\rangle+\langle{v_\theta^2}\rangle+\langle{v_\varphi^2}\rangle\right]r^2
dr,\label{27}
\end{eqnarray}
equation (\ref{26}) is reduced to
\begin{eqnarray}
2k-\frac{1}{2}\int^{R}_{0}4\pi
r^3\rho\left[\langle{v_t^2}\rangle+\langle{v_r^2}\rangle\right]\frac{\partial
\mu}{\partial r}dr=0.\label{28}
\end{eqnarray}
Let us write the energy-momentum tensor of the confined matter on
the brane in terms of the distribution function
\begin{eqnarray}
\tau_{\mu\nu}=\int f m v_{\mu}v_{\nu} dv,\label{29}
\end{eqnarray}
which gives
\begin{eqnarray}
\rho_{eff}=\rho\langle{v_{t}^{2}}\rangle,\hspace{.5
cm}p_{eff}^{(r)}=\rho\langle{v_{r}^{2}}\rangle,\hspace{.5
cm}p_{eff}^{(\perp)}=\rho\langle{v_{\theta}^{2}}\rangle=\rho\langle{v_\varphi^{2}}\rangle.\label{30}
\end{eqnarray}
Now, using these relations and summing the gravitational field
equations (\ref{20})-(\ref{22}) we find
\begin{eqnarray}
e^{-\nu}\left(\frac{\mu^{'}}{r}-\frac{\mu^{'}\nu^{'}}{4}+\frac{\mu^{''}}{2}+
\frac{\mu^{'2}}{4}\right)=4\pi
G\rho\langle{v^2}\rangle-\lambda+\frac{1}{\varepsilon}\left(-3\alpha^2-\frac{2\alpha\beta}{r}+
3\alpha ce^{-\mu/2}+\frac{2\beta c}{r}e^{-\mu/2}\right),\label{31}
\end{eqnarray}
where
$\langle{v^2}\rangle=\langle{v_t^2}\rangle+\langle{v_r^2}\rangle+
\langle{v_\theta^2}\rangle+\langle{v_\varphi^2}\rangle$. In order
to obtain the generalized virial theorem in our model we have to
use some approximations and assumptions. Since the dispersion of
the velocity of galaxies in the clusters is of the order $600-1000
$km/s, that is $(\frac{v}{c})^2\approx 4\times10^{-6} -
1.11\times10^{-5} \ll 1$, therefore we can neglect the
relativistic effects in the Boltzmann equation, and use the small
velocity limit approximation, so that
$\langle{v_r^2}\rangle\approx \langle{v_\theta^2}\rangle\approx
\langle{v_\varphi^2}\rangle\ll\langle{v_t^2}\rangle\approx1$. The
intensity of the gravitational effects can be estimated from the
ratio of the Schwarzschild radius to the radius of cluster,
$GM/R$, which for typical clusters is of the order of $10^{-6}\ll
1$. Therefore the gravitational field inside galactic clusters is
weak and we can use the weak gravitational field approximation, so
that the quadratic terms can be neglected \cite{Harko}. Thus,
equation (\ref{31}) is given by
\begin{eqnarray}
\frac{1}{2r^2}\frac{\partial}{\partial r}\left(r^2\frac{\partial
\mu}{\partial r}\right)=4\pi G\rho-\lambda+4\pi
G\rho_{extr},\label{32}
\end{eqnarray}
where
\begin{eqnarray}
4\pi
G\rho_{extr}=\frac{1}{\varepsilon}\left(-3\alpha^2-\frac{2\alpha\beta}{r}+3\alpha
ce^{-\mu/2}+\frac{2\beta c}{r}e^{-\mu/2}\right),\label{rho}
\end{eqnarray}
and the suffix `extr' stands for `extrinsic'. Multiplying equation
(\ref{32}) by $r^2$ and integrating from 0 to $r$ we obtain
\begin{eqnarray}
\frac{1}{2}r^2\frac{\partial \mu}{\partial r}-GM(r)+\frac{1
}{3}\lambda r^3-4\pi
G\int^r_0\rho_{extr}(r{'})r^{'2}dr{'}=0,\label{33}
\end{eqnarray}
where $M(r)$ is the mass out to radius $r$, so that
$dM(r)=4\pi\rho r^2dr$. The total baryonic mass of the system is
given by $M=\int^r_04\pi\rho r^{'2}dr{'}$. We also define the
geometrical mass as
\begin{eqnarray}
M_{extr}(r)=4\pi\int^r_0\rho_{extr}(r{'})r^{'2}dr{'}.\label{mass}
\end{eqnarray}
Multiplying equation (\ref{33}) by $dM(r)$ and integrating from 0
to $R$ and introducing the moment of inertia of the system as
$I=\int^R_0r^2dM(r)$, we obtain
\begin{eqnarray}
2k+W+\frac{1}{3}\lambda I-W_{extr}=0.\label{virial}
\end{eqnarray}
In the above equation we have used equation (\ref{27}) and the
following definitions
\begin{eqnarray}
W=-\int^R_0\frac{GM(r)}{r} dM(r),\label{35}
\end{eqnarray}
and
\begin{eqnarray}
W_{extr}=\int^R_0\frac{GM_{extr}(r)}{r} dM(r),\label{36}
\end{eqnarray}
where $W$ is the gravitational potential energy of the system. For
the case $\alpha=\beta=c=0$, we obtain $W_{extr}=0$ and equation
(\ref{virial}) reduces to the virial theorem in the presence of a
cosmological constant \cite{Jackson}. Now, let us introduce the
radii $R_v$ , $R_I$ and $R_{extr}$ as \cite{Harko}
\begin{eqnarray}
R_v=\frac{M^2}{\int_0^R \frac{M(r)dM(r)}{r}},
\end{eqnarray}
\begin{eqnarray}
R_I^2=\frac{\int_0^R r^2 dM(r)}{M(r)},
\end{eqnarray}
\begin{eqnarray}
R_{extr}=\frac{M^2_{extr}}{\int_0^R \frac{M_{extr}(r)dM(r)}{r}},
\end{eqnarray}
where $R_v$ is the virial radius and $R_{extr}$ is defined as the
geometrical radius of the clusters of galaxies. Defining the
virial mass as \cite{Jackson}
\begin{eqnarray}
2k=\frac{GM^2_v}{R_v},
\end{eqnarray}
and using the following relations
\begin{eqnarray}
W=-\frac{GM^2}{R_v},\hspace{.5
cm}W_{extr}=\frac{GM_{extr}^2}{R_{extr}},\hspace{.5
cm}I=MR_I^2,\label{37}
\end{eqnarray}
the generalized virial theorem (\ref{virial}) is simplified as
\begin{eqnarray}
\frac{M_v}{M}=\left(1-\frac{\lambda
R_vR_{I}^2}{3GM}+\frac{M^2_{extr}R_v}{M^2R_{extr}}\right)^{1/2}.\label{38}
\end{eqnarray}
As can be seen, we have three type of mass in equation (\ref{38}),
namely, the total baryonic mass of the system represented by $M$
(including the baryonic mass of the intra-cluster gas and of the
stars, other particles like massive neutrinos), the virial mass
represented by $M_v$ and finally, the geometrical mass represented
by $M_{extr}$.

On large distance scales associated with galaxies, we can ignore
the contribution of the cosmological constant relative to the mass
energy of the galaxy. Since, as it turns out, $M_{v}$ is
considerably greater than $M$ for most of the clusters, we can
neglect the unitary term in equation (\ref{38}). Therefore, the
virial mass is given by
\begin{eqnarray}
M_v\approx{M_{extr}}\left(\frac{R_{v}}{R_{extr}}\right)^{1/2}.\label{Virial}
\end{eqnarray}
This equation shows that the virial mass is proportional to the
geometrical mass due to the local extra term, $Q_{\mu\nu}$, thus
the virial mass discrepancy in clusters of galaxies can be
explained by the generalized virial theorem. Since galactic
clusters are dark matter dominated objects, the main contribution
to their mass comes from the geometrical mass $M_{extr}$ so that
with a very good approximation $M_{extr}\approx M_{total}$.

All the way, among various possible solutions that depend on
different choices of arbitrary constants $\alpha, \beta, c$, we
consider $\alpha=0$ and $\beta,c\neq0$. Since the bulk of matter
in clusters of galaxies is in the form of dark matter, we can
neglect the effect of the ordinary matter in equations
(\ref{20})-(\ref{22}). Thus, the solutions are given by
\begin{eqnarray}
e^{-\nu}=\left(1-\frac{\beta^2}{\varepsilon}\right),\label{40}
\end{eqnarray}
\begin{eqnarray}
e^{-\mu/2}=\frac{\left(\varepsilon-\beta^2\right)}{\beta
c}\frac{1}{r}.\label{41}
\end{eqnarray}
Now, using equations (\ref{rho}) and (\ref{mass}) the geometrical
mass is given by
\begin{eqnarray}
GM_{extr}(r)=\frac{1}{\varepsilon}\int^r_0\left[-3\alpha^2-\frac{2\alpha\beta}{r^{'}}
+3\alpha ce^{-\mu(r^{'})/2}+\frac{2\beta
c}{r^{'}}e^{-\mu(r^{'})/2}\right]r^{'2}dr^{'}.
\end{eqnarray}
Substituting $e^{-\mu/2}$ from equation (\ref{41}), we obtain
\begin{eqnarray}
GM_{extr}(r)=2\left(1-\frac{\beta^2}{\varepsilon}\right)r.\label{43}
\end{eqnarray}
Since according to our physical interpretation $M_{extr}$ is an
effective geometrical mass it must satisfy the condition
$M_{extr}>0$, therefore we take $\varepsilon=-1$ ($AdS_5$ bulk).
This geometrical mass is linearly increasing with $r$, thus having
a similar behavior as that of dark matter in clusters of galaxies.
An important observational quantity is the radial velocity
dispersion $\sigma_r^2$, which is related to the total mass in a
cluster by the relation $GM_{v}(R_v)=\sigma_r^2R_v$. For $r = R_v$
equation (\ref{43}) becomes $GM_{extr}(R_v)\approx(1+\beta^2)R_v$
and assuming that $M_{extr}\approx M_v$ we obtain
\begin{eqnarray}
(1+\beta^2)\approx\sigma_{r}^2.\label{44}
\end{eqnarray}
Therefore, the present model gives a geometrical interpretation
for the mass discrepancy in clusters of galaxies, supporting our
previous results that the galaxy rotation curves can be explained
without assuming the existence of dark matter and new modified
theories (Modified Newtonian Dynamics) \cite{Razmi}.

Finally, we should mention here that there is a difference between
our model and the work presented in \cite{Harko}. In this paper we
have shown that the virial mass of galaxy clusters is mainly
determined by the geometrical mass associated with the extrinsic
curvature terms (local bulk effects) whereas in \cite{Harko}, the
virial mass is specified by the dark mass associated with the
transmitted projection of the bulk Weyl tensor (non-local bulk
effects).
\section{Conclusions}
In the RS brane-world models, the 4-dimensional
effective Einstein equation has some extra terms known as dark
radiation and dark pressure, which describe the non-local bulk
effects due to the gravitational field of the bulk. It has been
shown that the dark radiation term can be used to explain the
virial theorem mass discrepancy in clusters of galaxies
\cite{Harko}.

In the present paper we have generalized the virial theorem within
the context of brane-world models without using the $Z_2$ symmetry
or without postulating any junction condition. To obtain the
virial theorem we have used a method based on the collisionless
Boltzmann equation. Assuming a constant curvature bulk, the virial
theorem was modified by a local extra term, equation
(\ref{virial}), which was then used to explain the virial mass
discrepancy in clusters of galaxies. Thus, the geometrical mass
can play the role of dark matter in clusters of galaxies,
supporting our previous results that the rotational galactic
curves can be explained in our model without introducing any
additional hypothesis \cite{Razmi}. Finally, after fixing the
parameter of this model, we found them to be in good agreement
with observations.

\end{document}